\documentclass[12pt]{article}
\def\nng{\nu \to \nu \gamma}
\def\prl{{\mbox{\tiny $\parallel$ }}}
\def\half{\frac{1}{2}} \def\lm{\lambda}
\def\ep{\varepsilon}
\usepackage{epsf}
\date{}

\title{Radiative neutrino transition $\nng$ in strongly magnetized
plasma}

\author{{M.V.~Chistyakov\thanks{E-mail address: mch@univ.uniyar.ac.ru}, N.V.~Mikheev\thanks{E-mail address: mikheev@yars.free.net}}\\ [7mm]
{\small\it Division of Theoretical Physics, Department of Physics,}\\
{\small\it Yaroslavl State University, Sovietskaya 14,}\\ {\small\it
150000 Yaroslavl, Russian Federation}}

\begin{document}
\oddsidemargin 5mm
\maketitle

\begin{abstract}
The influence of strongly magnetized electron-positron plasma on
the radiative neutrino transition $\nng$ is investigated. The
probability and mean losses of the neutrino energy and momentum
are calculated taking account of the photon dispersion and large
radiative corrections near the resonance. It is shown that the
combined effect of plasma and strong magnetic field decreases the
probability and mean values of the neutrino energy and momentum
loss in comparison with these values obtained in pure magnetic
field.

\medskip

\end{abstract}

\newpage

It is well known that neutrino processes play an important role in
astrophysics \cite{Raffelt}. In many cases these processes occur
in the presence of plasma and/or magnetic field. Of particular
conceptual interest are those processes which are forbidden or
strongly suppressed in vacuum. One such reaction is the radiative
massless neutrino transition $\nng$. Previously this process was
studied in plasma and magnetic field separately. In plasma the
process $\nng$ was firstly investigated in \cite{Tsyt} and more
recently in \cite{ONP,Mel}. In pure magnetic field the radiative
neutrino transition $\nng$ was studied in the papers
\cite{GN, Skob, IR, GMV}. In the framework of four-fermion theory
the amplitude and the probability of the process were calculated
in refs. \cite{GN} and \cite{Skob} in the crossed field and strong
magnetic field respectively. In the Standard Model the amplitude of
the neutrino transition $\nng$ was found in \cite{IR, GMV} for the
arbitrary magnetic field strength. In the paper \cite{IR} the case
of the moderate neutrino energies, $E_{\nu} < 2 m_e$ was studied
in the kinematical region where the final photon
dispersion law was closed to the vacuum one,  $q^2=0$. The limit of the
large neutrino energies and strong magnetic field was investigated
in \cite{GMV}. There is that case which could be realized at the
Kelvin-Helmholz stage of
supernova remnant cooling, when the energies of the neutrino are
$E_{\nu} \simeq 10-20$ MeV and the magnetic field strength could
be as high as $10^{16} - 10^{17}$ G \cite{BK}. It was shown in \cite{GMV}
that the main contribution into the probability of the
neutrino transition $\nng$ was determined from the vicinity of the
lowest cyclotron resonance, when the amplitude of the process and
photon polarization operator contained simultaneously the
square-root singularity.

The purpose of our work is to study the influence of the electron-positron
plasma on the process of the radiative massless neutrino transition $\nng$ in
a strong magnetic field. This process is considered in the framework of the
Standard Model using the effective local Lagrangian of the neutrino-electron
interaction
\begin{equation}
{\cal L} \, = \, \frac{G_F}{\sqrt 2}
\big [ \bar e \gamma_{\alpha} (g_V - g_A \gamma_5) e \big ] \,
j_{\alpha} \,,
\label{L}
\end{equation}
\noindent where $g_V = \pm 1/2 + 2 \sin^2 \theta_W, \, g_A = \pm 1/2$. Here
the upper signs correspond to the electron neutrino
($\nu = \nu_e$) when both $Z$ and $W$ boson exchange takes part
in a process. The lower signs correspond to $\mu$ and $\tau$ neutrinos ($\nu
= \nu_{\mu}, \nu_{\tau}$), when the $Z$ boson exchange
is only presented in the Lagrangian~(\ref{L}),
$j_{\alpha} = \bar \nu \gamma_{\alpha} (1 - \gamma_5) \nu$ is the left neutrino
current. We investigate the limit of ultrarelativistic
strongly magnetized plasma, when the magnetic field strength is the largest
physical parameter
\begin{equation}
eB > E_{\nu}^2, \, \mu^2, \, T^2 \gg m_e^2.
 \label{Cond}
\end{equation}
Here $\mu$ is the electron chemical potential, $T$ is the
temperature of plasma. Under these conditions electrons and
positrons in plasma occupy dominantly the lowest Landau level.

Notice that the amplitude and the probability of the process
$\nng$ depend essentially on the polarization of the final photon.
In a general case there exist three eigenmodes of the photon
polarization operator. The corresponding eigenvectors can be written
in the following form:
\begin{equation}
\varepsilon _{\mu}^{(1)} = \frac{ (q \varphi)_{\mu} } { \sqrt{
q^2_\perp  } } ; \; \; \; \; \; \varepsilon _{\mu}^{(2)} = \frac{
(q \tilde \varphi)_{\mu} }{ \sqrt{ q^2_{\mbox{\tiny $\parallel$ }}
}}; \; \; \; \; \; \varepsilon _{\mu}^{(3)} = \frac{q^2 (q
\varphi\varphi)_{\mu} - q_{\mu} (q \varphi\varphi q)} {\sqrt{q^2
q^2_{\mbox{\tiny $\parallel$ }} q^2_\perp }},
 \label{eps}
\end{equation}
where $ \varphi_{\alpha \rho} = F_{\alpha \rho} / B$ is the
dimensionless tensor of the external magnetic field, ${\tilde
\varphi}_{\alpha \rho} = \frac{1}{2} \varepsilon_{\alpha \rho \mu
\nu} \varphi_{\mu \nu} \; $ is the dual tensor, $q^2_{\parallel} =
( q \tilde \varphi \tilde \varphi q ) = q_\alpha \tilde
\varphi_{\alpha\rho} \tilde \varphi_{\rho\mu} q_\mu$, $q^2_{\perp}
=  ( q \varphi \varphi q )$. Only two of these modes,
$\ep^{(1)}_{\mu}$ and $\ep^{(2)}_{\mu}$ are the physical one
in the pure magnetic field.
\begin{footnote}
{The modes $\ep^{(1)}_{\mu}$ and $\ep^{(2)}_{\mu}$ correspond to
Adler's so-called parallel ($\|$) and perpendicular ($\bot$) modes
\cite{Adl}.}
\end{footnote}
As the analysis shows the presence of the strongly magnetized
plasma doesn't modify the eigenvectors (\ref{eps}) but modifies
the eigenvalue corresponding to the vector $\ep^{(2)}_{\mu}$ only.
This is due to the fact that the interaction of the two other
eigenmodes with the electrons and positrons which occupy the lowest Landau
level is strongly suppressed under the condition (\ref{Cond}). Hence,
only the photon with eigenvector $\ep^{(2)}_{\mu}$ can be created
in the process under consideration, as it takes place in the pure
magnetic field \cite{GMV}.

The process of the radiative neutrino transition is depicted by the
Feynman diagram, see Fig. 1, where the double line corresponds to
the propagator of electron in the presence of a magnetic field and
plasma. Several methods are known in literature describing the
process in the background plasma. In the present paper we use the
Real Time Formalism (RTF). The general expression of the real-time
propagator in an external field can be found in the paper
\cite{EPS}. In the limit of a strong magnetic field the propagator
can be presented in the form:
\begin{eqnarray}
S(x,y) = e^{i \Phi(x,y)} \int \frac{d^4 p}{(2 \pi)^4}\, S(p)\,
e^{-i p (x-y)}\label{S}
\end{eqnarray}
where
\begin{eqnarray}
&&S(p) \simeq 2 (\gamma p_\prl + m) \Pi_{-}
e^{-\frac{p_{\mbox{\tiny $\bot$}}^2}{eB}} \left (\frac{1}{p_\prl^2
- m^2 + i \varepsilon} - 2 i \pi f_F(p_0) \; \delta(p_\prl^2 - m^2)\right),
\\ [3mm]
&&f_F(p_0) = f_-(p_0) \theta(p_0) + f_+(-p_0)
\theta(-p_0),\qquad \Pi_- = \frac{1}{2} (1- i \gamma_1
\gamma_2).\nonumber
\end{eqnarray}
In (\ref{S}) $f_{\mp}(p_0)$ are the distribution functions of electron
and positron in plasma
\begin{eqnarray}
f_{\mp}(p_0) = \frac{1}{e^{\frac{p_0 \mp \mu}{T}} + 1}. \nonumber
\end{eqnarray}
Notice that in the case of two-point function the translationally
and gauge noninvariant phase factors $\Phi(x,y)$ are cancelled:
$$\Phi(x,y)+\Phi(y,x)=0.$$ Using the propagator (\ref{S}) the
amplitude of the process can be presented in the form:
\begin{eqnarray}
{\cal M} = {\cal M}_B + {\cal M}_{pl},
\label{Mgen}
\end{eqnarray}
where ${\cal M}_B$ is the amplitude of the process $\nng$
corresponding to the pure magnetic field contribution ($T = \mu =
0$). Following \cite{GMV} it can be expressed in the form
 \begin{footnote}
{Hereafter all expressions contain the scalar production of
4-vectors in Minkowski subspace (0, 3) only (field {\bf B} is
directed along the third axis). The metric of the subspace is
defined by $g_{\|} = \tilde \varphi \tilde \varphi = (+, 0, 0,
-)$. Therefore, for arbitrary 4-vectors $a_{\mu}$, $b_{\mu}$ one
has $(ab) =(a \tilde \varphi \tilde \varphi b) = a_0 b_0 - a_3
b_3$.}
\end{footnote}
\begin{eqnarray}
&&{\cal M}_B = \frac{e G_F} {2 \pi^2
\sqrt{2}}\,\,\frac{eB}{\sqrt{q^2}} \,\{ g_V (j\tilde \varphi q) +
g_A (jq) \}\, H \left ( \frac{4 m_e^2}{q^2} \right ),
\nonumber \\
\label{M_B}\\
&&H(z)=\frac{z}{\sqrt{z - 1}} \arctan \frac{1}{\sqrt{z - 1}} - 1,
\ z > 1,
\nonumber \\
&&H(z) = - \half \left ( \frac{z}{\sqrt{1-z}}
\ln \frac{1 + \sqrt{1-z}}{1 - \sqrt{1-z}} + 2 +
i \pi \frac{z}{\sqrt{1-z}} \right ), \ z < 1.
\nonumber
\end{eqnarray}
It should be noted that ${\cal M}_B$ is the amplitude with the
definite photon polarization corresponding to the mode
$\ep^{(2)}_{\mu}$ from the equation (\ref{eps}).

 The second term in (\ref{Mgen}), ${\cal M}_{pl}$, is induced by the
coherent neutrino scattering on plasma electrons and positrons with
photon radiation. For ${\cal M}_{pl}$ we find
\begin{eqnarray}
{\cal M}_{pl} = - \frac{e G_F}{\pi^2 \sqrt{2}} (eB)
m_e^2 \sqrt{q^2} \,
\{ g_V (j\tilde \varphi q) + g_A (jq) \}\,
\int \frac{d p_z}{E} \frac{f_{-}(E)+f_{+}(E)}
{4 (pq)^2 - (q^2)^2}.
\label{M_pl}
\end{eqnarray}

As was mentioned above the amplitude ${\cal M}_B$ contains the square-
root singularity which is connected with the cyclotron resonance on
the lowest Landau level. In the vicinity of the resonance point $q^2 = 4
m^2_e$ it becomes:
\begin{eqnarray}
{\cal M}_B \simeq \frac{e G_F}
{4 \pi \sqrt{2}}\,\,
\frac{eB}{\sqrt{4 m_e^2 - q^2}}
\,\{ g_V (j\tilde \varphi q) + g_A (jq) \}.
\end{eqnarray}
%
It is particularly remarkable that the amplitude ${\cal M}_{pl}$
contains the singularity of the same type.
In the limit $q^2 \to 4 m^2_e$ the total amplitude (\ref{Mgen})
can be presented in the following form:
\begin{eqnarray}
{\cal M} \simeq {\cal M}_B \; {\cal F} (q_0),
\label{M}
\end{eqnarray}
\noindent where
\begin{eqnarray}
{\cal F}(q_0) = \frac{\sinh{x}}{\cosh{x} + \cosh{\eta}},\qquad
x   = \frac{\vert q_0 \vert}{2 T}, \, \, \eta =
\frac{\mu}{T}.\nonumber
\end{eqnarray}

It should be stressed that not only the amplitude ${\cal M}$ has the
singular behaviour but the photon polarization ${\cal P}^{(2)}$ as
well. It can be obtained from (\ref{M}) by the following replacements
$$
{\cal P}^{(2)} =
- {\cal M} (\frac{G_F}{\sqrt{2}} g_V \to  e,
g_A \to 0, j_{\alpha} \to \varepsilon_{\alpha}^{(2)}).
$$
For ${\cal P}^{(2)}$ one has:
\begin{eqnarray}
{\cal P}^{(2)} \simeq - \frac{2 \alpha eB m_e}
{\sqrt{4 m_e^2 - q^2}}\; {\cal F} (q_0).
\end{eqnarray}
A large value of ${\cal P}^{(2)}$ near the resonance requires taking
account of large radiative corrections which reduce to a renormalization
of the photon wave function:
\begin{eqnarray}
\varepsilon_{\alpha}^{(2)} \to \varepsilon_{\alpha}^{(2)} \sqrt{Z}, \quad
Z^{-1} = 1 - \frac{\partial {\cal P}^{(2)}}{\partial q^2}.
\label{Z}
\end{eqnarray}
Using the formula (\ref{Z}) for the  amplitude we find
\begin{eqnarray}
{\cal M} \to \sqrt{Z}{\cal M} \simeq \frac{e G_F}{4 \pi}\,
\frac{eB}{\sqrt{q_{\mbox{\tiny $\bot$}}^2}} \,
\{ g_V (j\tilde \varphi q) + g_A (jq) \}{\cal F} (q_0). \label{M_end}
\end{eqnarray}
Thus, the photon wave-function renormalization
corrects the singular behaviour of the amplitude.

The probability of the process $\nng$ can be obtained by integration
of the amplitude over the phase space with taking account of the
photon dispersion \\
$q^2 - q_{\bot}^2 = {\cal P}^{(2)}$.
\begin{eqnarray}
E_{\nu} W &=& \frac{1}{32 \pi^2} \int \vert {\cal M} \sqrt{Z}
\vert^2
\, \delta(E_{\nu} - E_{\nu'} -
q_0({\bf k} - {\bf k'})) \times \nonumber \\
&\times& \frac{1}{1 - e^{- q_0/T}} \cdot
\frac{d^3 k'}{E_{\nu'} q_0}
\label{defW}
\end{eqnarray}
We assume that the neutrino distribution is closed to the Boltzman
one, so one can neglect the deviation of the neutrino statistical
factor from the unity. The probability (\ref{defW}) is rather
complicated in the general case and it will be published in the
extended paper. Here we present the results of our calculation in
two limiting cases of the cold, $\mu \gg T$, and hot, $T \gg \mu$,
plasma. Notice that in the vicinity of the cyclotron resonance,
which gives the main contribution to the probability, the photon
dispersion has a rather simple form $q_0 \simeq \sqrt{q^2_3 + 4
m_e^2}$. In the limit of the low temperature, $\mu \gg T$, for the
probability we obtain:
%
\begin{eqnarray}
W_{LT} &\simeq& \frac{\alpha (G_F eB)^2}{16 \pi^2} E_{\nu} \bigg \{
(g_V-g_A)^2
[(1 -\lm^2) - \frac{4 \mu}{E_{\nu}}(1 + \lm)] 
\theta(1 - \lm - \frac{4 \mu}{E_{\nu}})
+ \nonumber \\
&+& \, (g_V+g_A)^2\,
[(1 -\lm^2) - \frac{4 \mu}{E_{\nu}}(1 - \lm)]
\theta(1 + \lm - \frac{4 \mu}{E_{\nu}}).
\bigg \}.
\label{WLT}
\end{eqnarray}
Here $\lm$ is the cosine of the angle between the initial neutrino
momentum ${\bf k}$ and the magnetic field direction. In the
opposite limit of high temperature, $T \gg \mu$, the result for the
probability of the process $\nng$ is:
\begin{eqnarray}
W_{HT} &\simeq& \frac{\alpha (G_F eB)^2}{4 \pi^2} T \bigg \{ (g_V
-  g_A)^2 (1 + \lm) \,  F_1\left (\frac{E_{\nu} (1 - \lm)}{8
T}\right) + \nonumber \\ &+& (g_V+g_A)^2 (1 - \lm)\,
F_1\left(\frac{E_{\nu} (1 + \lm)}{8 T}\right) \bigg \},
\label{WHT}\\ [3mm]
 F_1(x) &=& x + \ln (\cosh x ) -
\frac{1}{4} \tanh^2 x - \tanh x \nonumber
\end{eqnarray}
In the limit of the rarefied plasma both expressions (\ref{WLT})
and (\ref{WHT}) reproduce the known formula for the radiative
neutrino transition probability in the pure strong magnetic field
\cite{GMV}:
\begin{eqnarray}
W_B \simeq \frac{\alpha (G_F eB)^2}{8 \pi^2 }(g^2_V + g^2_A)
E_{\nu} (1 - \lm^2). \label{WB}
\end{eqnarray}
In the case of the electron neutrino $\nu_e$ the dependence of the
probabilities (\ref{WLT}), (\ref{WHT}), (\ref{WB}) on the initial neutrino
energy are presented in Fig. 2, Fig. 3.

Keeping in mind a possible application of our results in
astrophysics we calculate the mean values of the neutrino energy
and momentum losses. These values could be defined by the
four-vector:
\[ Q_{\mu} = E_{\nu} \int d W \cdot q_{\mu} = - E_{\nu} (\frac{d
E_{\nu}}{d t}, \frac{d {\bf P}_{\nu}}{d t}), \]
where zero component $Q_0$ of this vector is connected with the
mean neutrino energy loss per unit time, the spatial components
${\bf Q}$ are connected with the momentum loss per unit time.

 For the zero and third components of the $Q_{\mu}$
 we obtain the following expression in the limit of cold plasma, $T \ll \mu$:
\begin{eqnarray}
Q_{0,3} &\simeq& \frac{\alpha (G_F eB)^2}{64 \pi^2} E_{\nu}^3 (1-
\lm^2) \bigg \{ (g_V + g_A)^2\,
[(1 + \lm) - \frac{16 \mu^2}{E_{\nu}^2 (1 + \lm)}] \,
\theta(1 + \lm - \frac{4 \mu}{E_{\nu}}) \nonumber \\ &\pm& \,
(g_V-g_A)^2\,
[(1 - \lm) - \frac{16 \mu^2}{E_{\nu}^2 (1 - \lm)}] \, \theta(1 -
\lm - \frac{4 \mu}{E_{\nu}}) \bigg \}. \label{QLT}
\end{eqnarray}
In the opposite case, when $\mu \ll T$ we find
\begin{eqnarray}
Q_{0,3} &\simeq& \frac{\alpha (G_F eB)^2}{2 \pi^2} E_{\nu} T^2
\bigg \{ (g_V + g_A)^2 (1 - \lm) \,  F_2\left (\frac{E_{\nu} (1 +
\lm)}{4 T}\right)  \nonumber \\ &\pm& (g_V+g_A)^2 (1 + \lm)\,
F_2\left(\frac{E_{\nu} (1 - \lm)}{4 T}\right) \bigg \},
\label{QHT}\\ [3mm] F_2(x) &=& \frac{1}{2}\tanh \frac{x}{2} -
\frac{x e^x(1+2e^x)}{(1 + e^x)^2} + (2+x) \ln (1 + e^x) + Li_2(-
e^x) -  \ln 4 + \frac{\pi^2}{12},\nonumber
\end{eqnarray}
where $Li_2(x)$ is the polylogarythm function. Notice that in the
limit $T \to 0, \, \mu \to 0$
both expressions (\ref{QLT}) and (\ref{QHT}) reproduce the formula
for the four-vector of losses in the pure strong magnetic field
\cite{GMV}:
\begin{eqnarray}
Q_{0,3} = \frac{\alpha (G_F eB)^2}{64 \pi^2} E_{\nu}^3 (1-\lm^2)
\bigg \{ (g_V + g_A)^2 (1 + \lm) \pm (g_V - g_A)^2 (1 - \lm) \bigg
\}. \label{QB}
\end{eqnarray}
We note, that electron-positron plasma and photon gas make an
opposite influence on the process under consideration. On one
hand,
the electron-positron background decreases
the amplitude of the process
(${\cal F}(q_0)< 1$). On the other hand, the probability and the mean
value of the neutrino energy and momentum loss increases due to
the effect of the stimulated photon emission. The analysis
shows that the combined effect of electron-positron plasma and
photon gas leads to the decreasing of the probability
in comparison to the result in the strong magnetic field
(\ref{WB}) as one can see from the Fig. 2, Fig. 3.
The similar supressing plasma influence on four-vector of neutrino energy
and momentum losses takes place.

In conclusion, we have calculated the process of the radiative
neutrino transition in the presence of plasma and strong magnetic
field. It was shown that the combined effect of plasma and strong
magnetic field decreases the probability and mean values of
neutrino energy and momentum loss
in comparison with these values obtained in pure magnetic field.
Therefore the complex medium plasma + strong magnetic field is
more transparent to neutrino with regard to the process $\nng$, than the pure
magnetic field.
\bigskip

\noindent {\bf Acknowledgments}
This work was supported
in part by the INTAS Grant N~96-0659, by the Russian Foundation
for Basic Research Grant N~98-02-16694 and by the International Soros
Science Education Program under the Grants N~a99-1432 (M.C.).

\newpage

\thispagestyle{empty}

\centerline{\bf Figure captions}

\centerline{to the paper: M.V.~Chistyakov, N.V.~Mikheev, ``Radiative neutrino transition \dots''}

\vspace{5mm}

\begin{itemize}

\item[\bf Fig. 1]
The Feynman diagram for the radiative neutrino transition in the presence
of plasma and magnetic field.

\item[\bf Fig. 2]
The dependence of the probability of the radiative neutrino transition
$\nu_e \to \nu_e \gamma$ on energy in strongly magnetized cold plasma, $T
\ll \mu$. The dotted lines correspond to $\nu_e \to \nu_e \gamma$ process
in pure magnetic field. The lines 1, 2, 3 depict the probabilities for
angles between the initial neutrino momentum and the magnetic field
direction $\theta = \pi/4, \pi/2, 3 \pi/4$ correspondingly. Here
$W_0 = \alpha (G_F eB)^2 \mu / 8 \pi^2$.

\item[\bf Fig. 3]
The probability $\nu_e \to \nu_e \gamma$ process in strongly magnetized
hot plasma, $T \gg \mu$. Here $W_0 = \alpha (G_F eB)^2 T / 8 \pi^2$,
other notations are the same as in the Fig. 2.

\end{itemize}

\newpage
\thispagestyle{empty}

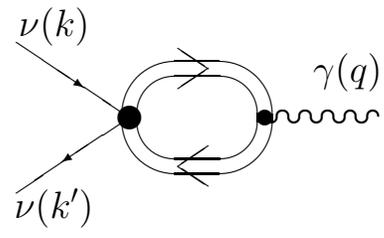
\begin{figure}[ht]

\def\photonatomright{\begin{picture}(3,1.5)(0,0) \put(0,-0.75){\tencircw
                                \symbol{2}} \put(1.5,-0.75){\tencircw
                                \symbol{1}} \put(1.5,0.75){\tencircw
                                \symbol{3}} \put(3,0.75){\tencircw
                                \symbol{0}}
                      \end{picture}
                     } \def\photonatomup{\begin{picture}(1.5,3)(0,0)
                             \put(-0.75,3){\tencircw \symbol{3}} \put(-
                             0.75,1.5){\tencircw \symbol{2}}
                             \put(0.75,1.5){\tencircw \symbol{0}}
                             \put(0.75,0){\tencircw \symbol{1}}
                   \end{picture}
                  } \def\photonrighthalf{\begin{picture}(30,1.5)(0,0)
                     \multiput(0,0)(3,0){5}{\photonatomright}
                  \end{picture}
                 }
\def\photonuphalf{\begin{picture}(1.5,15)(0,0)
                      \multiput(0,0)(0,3){5}{\photonatomup}
                   \end{picture}
                  }



\unitlength=1.00mm
\special{em:linewidth 0.4pt}
\linethickness{0.4pt}

\begin{picture}(40.00,45.00)(-25.00,0.00)
\put(35.00,32.50){\oval(20.00,15.00)[]}
\put(35.00,32.50){\oval(16.00,11.00)[]}
\put(26.00,32.50){\circle*{3.00}}
\put(44.00,32.50){\circle*{2.00}}
\linethickness{0.8pt}
\put(11.00,42.50){\vector(3,-2){9.00}} \put(26.00,32.50){\line(-
3,2){6.00}} \put(26.00,32.50){\vector(-3,-2){9.00}}
\put(17.00,26.50){\line(-3,-2){6.00}}
\put(36.50,39.00){\line(-3,2){4.01}} \put(36.50,39.00){\line(-3,-
2){4.01}}
\put(32.50,26.00){\line(3,2){4.01}} \put(32.50,26.00){\line(3,-
2){4.01}}
%
\put(16.00,42.00){\makebox(0,0)[cb]{\large $\nu(k)$}}
\put(16.00,23.00){\makebox(0,0)[ct]{\large $\nu(k')$}}
\put(55.00,36.00){\makebox(0,0)[cb]{\large $\gamma(q)$}}
%
%
\def\photonatomright{\begin{picture}(3,1.5)(0,0) \put(0,-
                                0.75){\tencircw \symbol{2}} \put(1.5,-
                                0.75){\tencircw \symbol{1}}
                                \put(1.5,0.75){\tencircw \symbol{3}}
                                \put(3,0.75){\tencircw \symbol{0}}
                      \end{picture}
                     }
\def\photonrighthalf{\begin{picture}(30,1.5)(0,0)
                     \multiput(0,0)(3,0){5}{\photonatomright}
                  \end{picture}
                 }
\put(44.00,32.50){\photonrighthalf}
\end{picture}



%
%
\caption{M.V.~Chistyakov, N.V.~Mikheev ``Radiative neutrino transition \dots''}
\end{figure}

\newpage
\thispagestyle{empty}

\begin{figure}[ht]
\epsffile[140 320 0 700]{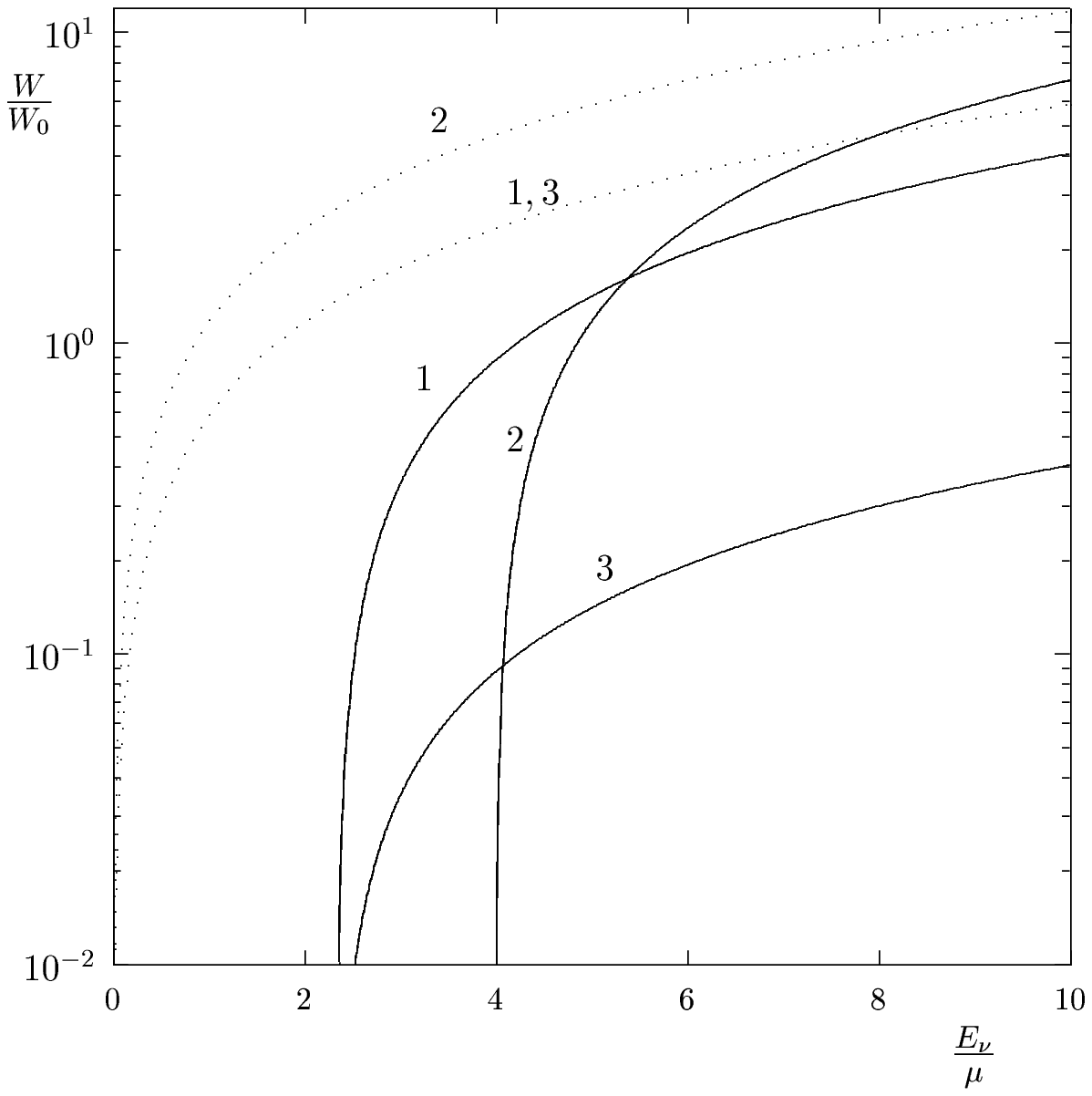}
\caption{M.V.~Chistyakov et al., ``Radiative neutrino \dots''}
\end{figure}

\newpage
\thispagestyle{empty}

\begin{figure}[ht]
\epsffile[140 320 0 700]{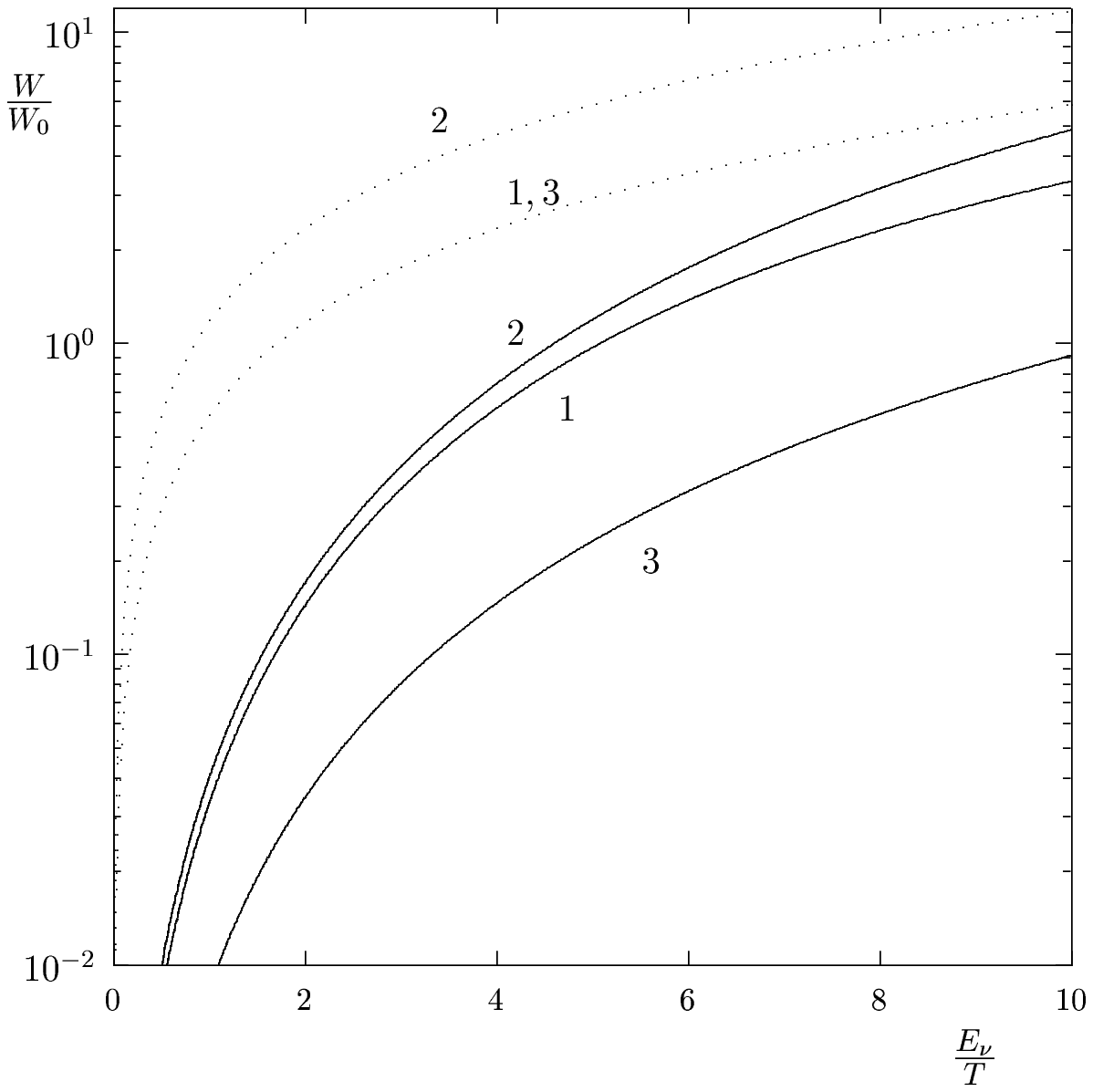}
\caption{M.V.~Chistyakov et al., ``Radiative neutrino \dots''}
\end{figure}


\begin{thebibliography}{99}
%
\bibitem{Raffelt}
   G.G.~Raffelt, Stars as Laboratories for Fundamental Physics,
     University of Chicago Press, Chicago, 1996.
%
\bibitem{Tsyt}
V.~N.~Tsytovich, J. Exptl. Theoret. Phys. (U.S.S.R)
{\bf 45}, (1963)1183  [Sov. Phys. JETP {\bf 18}, 816 (1964)].
%
\bibitem{ONP}
  J.~D'Olivo, J.~Nieves, P.~Pal,
  Phys.Lett.B {\bf 365} (1996) 178.
%
\bibitem{Mel}
S.~J.~Hardy and D.~B.~Melrose, Publ. Astron. Soc. Aus. {\bf
13}, (1996) 144.
%
\bibitem{GN}
    D.~V.~Galtsov, N.~S.~Nikitina,
    Zh. Eksp. Theor. Fiz. 62 (1972) 2008.
%
\bibitem{Skob}
    V.V.~Skobelev, Zh. Eksp. Teor. Fiz. 71 (1976) 1263
    [Sov. Phys. JETP 44 (1976) 660].
%
\bibitem{IR}
      A.N.~Ioannisian and G.G.~Raffelt, Phys.Rev. D55 (1997) 7038.
%
\bibitem{GMV}
   A.A.~Gvozdev, N.V.~Mikheev and L.A.~Vassilevskaya, Phys.~Lett. B410
   (1997) 211; In: Proc. Int. Seminar ``Quarks-96'' (Yaroslavl, Russia,
   1996).
%
\bibitem{BK}
  G.S.~Bisnovatyi-Kogan and S.G.~Moiseenko, Astron. Zh., 69 (1992) 563
   [Sov. Astron., 36 (1992) 285];
   G.S.~Bisnovatyi-Kogan, Astron. Astrophys. Transactions, 3 (1993) 287.
%
\bibitem{Adl}
   S.L.~Adler, Ann. Phys. N.Y. 67 (1971) 599.
%
\bibitem{EPS}
P.~Elmfors, D.~Persson and B.~Skagerstam, Nucl. Phys. B 464 (1996) 153.
\end{thebibliography}
\end{document}